\documentclass[aip,jcp]{revtex4-1}

\usepackage{graphicx}
\usepackage{color}
\usepackage[welsh,english]{babel}

\draft 

\begin{document}


\title{Dynamics of hard sphere suspensions using Dynamic Light Scattering and X-ray  Photon Correlation Spectroscopy: dynamics and scaling of the Intermediate Scattering Function.} 



\author{V.A. Martinez}
\email[]{vincent.martinez@ed.ac.uk}
\affiliation{SUPA, COSMIC and School of Physics and Astronomy, The University of Edinburgh, King's Buildings, Mayfield Road, Edinburgh EH9 3JZ, United Kingdom}
\affiliation{Department of Applied Physics, Royal Melbourne Institute of Technology, Melbourne, VIC 3000, Australia}
\author{J.H.J. Thijssen}
\affiliation{SUPA, COSMIC and School of Physics and Astronomy, The University of Edinburgh, Kings Buildings, Mayfield Road, Edinburgh EH9 3JZ, United Kingdom}
\author{F. Zontone}
\affiliation{European Synchrotron Radiation Facility (ESRF), BP 220, F-38043 Grenoble Cedex,France}
\author{W. van Megen}
\author{G. Bryant}
\affiliation{Department of Applied Physics, Royal Melbourne Institute of Technology,Melbourne, VIC 3000, Australia}


\date{\today}

\begin{abstract}
Intermediate Scattering Functions (ISF's) are measured for colloidal hard sphere systems using both Dynamic Light Scattering (DLS) and X-ray Photon Correlation Spectroscopy (XPCS). We compare the techniques, and discuss the advantages and disadvantages of each. Both techniques agree in the overlapping range of scattering vectors. We investigate the scaling behaviour found by Segre and Pusey \cite{Segre1996} but challenged by Lurio et al. \cite{Lurio2000}. We observe a scaling behaviour over several decades in time but not in the long time regime. Moreover, we do not observe long time diffusive regimes at scattering vectors away from the peak of the structure factor and so question the existence of a long time diffusion coefficients at these scattering vectors. 
\end{abstract}


\maketitle 

\section{Introduction}
Over the past decade, X-ray Photon Correlation Spectroscopy (XPCS) has emerged as a valuable tool for the study of dynamics in soft matter \cite{Robert2005,Grubel2004,Thurn-Albrecht2003,Tsui1998,Grubel2000}. The technique has two principle advantages over the well established technique of Dynamic Light Scattering (DLS): the first is that the shorter wavelengths allow for measurements at larger scattering vectors (smaller size scales); the second is that X-ray measurements do not suffer from multiple scattering. These advantages are balanced against two major disadvantages:  X-rays have lower coherence than (laser) light sources (leading to reduced correlation intercepts); and that samples can be easily damaged by intense X-rays.

A number of authors have applied XPCS to study the dynamics of suspensions of colloidal particles \cite{Lurio2000,Grubel2000,Bandyopadhyay2004,Robert2008,Autenrieth2007,Pontoni2003}. This builds on extensive studies using light scattering techniques \cite{Megen1998,Megen2009}, and, more recently, microscopy techniques \cite{Prasad2007,Lynch2008,Kegel2000}. The most widely studied system is that of particles which behave as hard spheres \cite{Pusey1986}.

The fundamental quantity measured in scattering experiments is the Intermediate Scattering Function (ISF) which describes the dynamics of the particle number density. For suspensions at moderate concentrations, the dynamics are often divided into three regimes: a short time diffusive regime, where particles diffuse within their neighbour cages; a non-diffusive crossover regime or plateau, where the interactions between a particle and its neighbour cage are most clearly exposed; and a long time diffusion regime where the particles have escaped their neighbour cages. This long time regime is particularly difficult to access experimentally, and ambiguities in interpretation can arise depending on the method of analysis used. 

To date there have been no direct comparisons between the two techniques on concentrated suspensions of colloidal particles. There are however instances where XPCS and DLS experiments have not been in agreement. Specifically, DLS experiments carried out on sterically stabilized hard spheres \cite{Segre1996}, found that the short and long time diffusion coefficients ($D_{s}(q)$ and $D_{L}(q)$ respectively) both scale with the structure factor \-- in other words the ratio $D_{s}(q)/D_{L}(q)$ is independent of q. However, this finding has since been questioned by Lurio et al. \cite{Lurio2000}, who used XPCS to study a system of charge-stabilized polystyrene latex spheres in glycerol. In this work they showed that the measured structure factors for this system were consistent with those of hard spheres \-- however, when they investigated the dynamics of this system, they did not observe the long time scaling. There are several possible explanations for this discrepancy: i) there is a fundamental difference between XPCS and DLS which results in different measured behaviours; ii) X-ray damage contributes to the measured results in XPCS; iii) the charge-stabilized pseudo-hard-sphere system is not equivalent to the steric hard sphere system; and/or iv) the differences may arise from ambiguities in the determination of the long time diffusion coefficient \cite{Williams2006,Megen2007}.
 
In this paper we compare the two techniques by examining sterically stabilized hard spheres using both DLS and XPCS. By using the same samples for both techniques we can eliminate sample variation and effects of sample preparation. We conduct analyses similar to those in Lurio et al. \cite{Lurio2000} and Segre and Pusey \cite{Segre1996}, and compare the results of the two methods. A previous study \cite{Riese2000} investigated the comparison between DLS and XPCS but only at a low volume fraction, 0.164. In that work the overlapping range of scattering vectors did not include the peak of the structure factor or the minimum of the form factor, where contributions from multiple scattering are maximal.

\section{Theory}
DLS and XPCS are now standard methods for measuring the dynamics of colloidal suspensions under certain conditions and have been reviewed in numerous articles\cite{Grubel2004,Grubel2000,Megen1998,Riese2000,Megen2007}. In this section, we summarize some important aspects. The basic quantity measured by both techniques is the (normalized) time-averaged autocorrelation function (ICF) of the intensity, I(q,t), of the scattered light:
 \begin{equation}
g^{(2)}(q,\tau)=\frac{\langle I(q,0)I(q,\tau)\rangle}{\langle I(q,0)\rangle^2}
\end{equation}
where $q$ is the magnitude of the scattering vector, $\tau$ is the delay time. For ergodic systems the ICF factors according to the Siegert relationship:
\begin{equation}\label{eq:Siegert}
g^{(2)}(q,\tau)=1+c\mid f(q,\tau)\mid^{2}
\end{equation}
where
\begin{equation}
f(q,\tau)=\frac{F(q,\tau)}{F(q,0)}
\end{equation}
and
\begin{equation}
F(q,\tau)=\frac{1}{N}\sum_{j,k=1}\langle \mathrm{exp}\left[i\vec{q}(\vec{r_{j}}(\tau)-\vec{r_{k}}(0))\right]\rangle
\end{equation}
is the intermediate scattering function \-- the auto-correlation function of the $q^{th}$ spatial Fourier component of the particle number density fluctuations.  The polydispersity of the colloidal suspension, i.e. the (small) relative spread in particle radii (see below), is not explicitly taken into account.  In Eq. (\ref{eq:Siegert}), c is an experimental constant determined by the ratio of the coherence area to the detector area.  Note also that $F(q,0)=S(q)$ is the static structure factor.

For diffusive density fluctuations we have,
\begin{equation}
f(q,\tau)=\mathrm{exp}(-q^{2}D(q)\tau)
\end{equation}
More generally one calculates  the time-dependent quantity \cite{Segre1996,Pusey1991}
\begin{equation}
D(q,\tau)=-\frac{1}{q^2}\frac{\mathrm{d}(\mathrm{ln}(F(q,\tau)))}{\mathrm{d}\tau}
\end{equation}
The latter describes the evolution from short time processes, expressing the particles' (diffusive) response to those hydrodynamic modes in the suspending liquid that propagate instantaneously on the experimental time scale $\tau>10^{-6}s$, to the long-time processes associated with structural rearrangement or diffusion over larger distances. These processes are characterised by the short and long time collective diffusion coefficients respectively:
\begin{equation}
\quad D_{s}(q)=\lim_{\tau\rightarrow \tau_l}\left[D(q,\tau)\right] \quad \text{and} \quad D_{L}(q)=\lim_{\tau\rightarrow \infty}\left[D(q,\tau)\right]
\end{equation}
Where $\tau_l$ is the lower limit of the experimental time window. It is now well established that $D_{s}(q)$ shows a minimum at the position, $q_m$, of the maximum in the static structure factor, $S(q)$ \cite{Segre1996}. This is the familiar de-Gennes narrowing \-- the initial diffusive decay of density fluctuations is slowest at $q_m$ \cite{Hansen1986, Megen1985}.

\section{Methods}
\subsection{Sample Preparation}
The particles used in this work consist of a co-polymer core of methymethacrylate (MMA) and Trifluorethylmetacrylate (TFEMA). The co-polymer TFEMA is added to enable refractive index matching in a single solvent (cis-decalin), and the suspension turbidity can be controlled by changing the temperature. For these samples the index matching is sufficiently good that multiple scattering is nearly zero in the range of scattering angles studied here. The particles are sterically stabilised by a thin coating of poly-12-hydroxystearic acid (PHSA), approximately 10 nm thick. The particles used here are designated XL63, and have a radius $R=185 nm$ and a polydispersity of $8-9\%$ \cite{Underwood1996,Bryant1999,Bryant2003}. All times are presented in units of the Brownian time, $\tau_B=R^2/6D_o=0.013s$ where $D_o$ is the diffusion constant for freely diffusing particles. The hard sphere-like interaction between these particles in these suspensions is confirmed by the identification of the equilibrium phase boundaries, i.e. the freezing and melting points. 

Samples for DLS were prepared in 8mm path length, sealable scattering glass cells. Samples for XPCS were transferred to quartz capillaries of 1.5mm diameter (Wolfgang Muller Glas Technik, Berlin). A ball bearing was placed into the cell to facilitate the tumbling of the particle suspension to ensure good mixing. The cells were sealed with araldite, ensuring evaporation was insignificant over timescales of months. For some test measurements (see below) DLS measurements were also carried out in the X-ray capillaries prior to or following the X-ray measurements.

An ALV-6010 spectrometer is used for the DLS measurements. A HeNe laser of wavelength 632.8 nm illuminates the sample, and an Avalanche Photon Detector (APD), located on a goniometer, measures the scattered intensity at a specific scattering angle $\theta$. The accessible angular range is $15-150^{\circ}$, corresponding to a q range of $\sim3.8-28 \mu m^{-1}$, encompassing the peak in the structure factor at $q\sim19 \mu m^{-1}$.

The XPCS experiments were performed at the ID10a station of the Troika beamline at the European Synchrotron Radiation Facility (ESRF) in Grenoble, France. A third generation synchrotron, combined with an undulator insertion device, provides a partially coherent X-ray beam which has sufficient brilliance for XPCS experiments. The standard "coherent" set-up is based on two factors: (1) a strong collimation of the beam to match the transverse coherence length $\xi_T$ of the TROIKA undulator source, $\xi_T=\lambda L/s$, where $s$ is the source size and $L$ the sample-to-source distance; and (2) wavelength selection using a standard Si(111) monochromator that provides the longitudinal coherence length $\xi_L=\lambda^{2}/\Delta \lambda$.  At the ID10A beamline this routinely yields a 10x10 $\mu m^2$ beamspot at the sample with a $1 \mu m$  longitudinal coherence length at 8 keV ($\lambda = 1.555 \AA$), well inside the limit for path length differences in the present range of scattering vectors $q$ \cite{Grubel2004}. Focusing optics (compound refractive lenses) can be inserted into the beam to optimise the intensity whenever the coherence conditions can be relaxed. In this configuration the "coherent" intensity on the sample can exceed $10^{10}$ photons/s. To match the speckle size a 0D detector (scintillation counter) is placed behind a 100 $\mu m$ aperture at a distance $D=2240m$m behind the sample. Measurements of the intensity fluctuations at small angles $\theta=d/D$ are achieved by translating the detector by a distance $d$ perpendicular to the incident beam. A digital autocorrelator, connected to the detector, allows the direct measurement of the ICF at a specific scattering vector of amplitude $q$ in the range $\sim10-100 \mu m^{-1}$. Capillaries were mounted in an evacuated sample chamber with temperature control ($T=21^{\circ}C$).

\section{Results}
\begin{figure}
\includegraphics[scale=0.4]{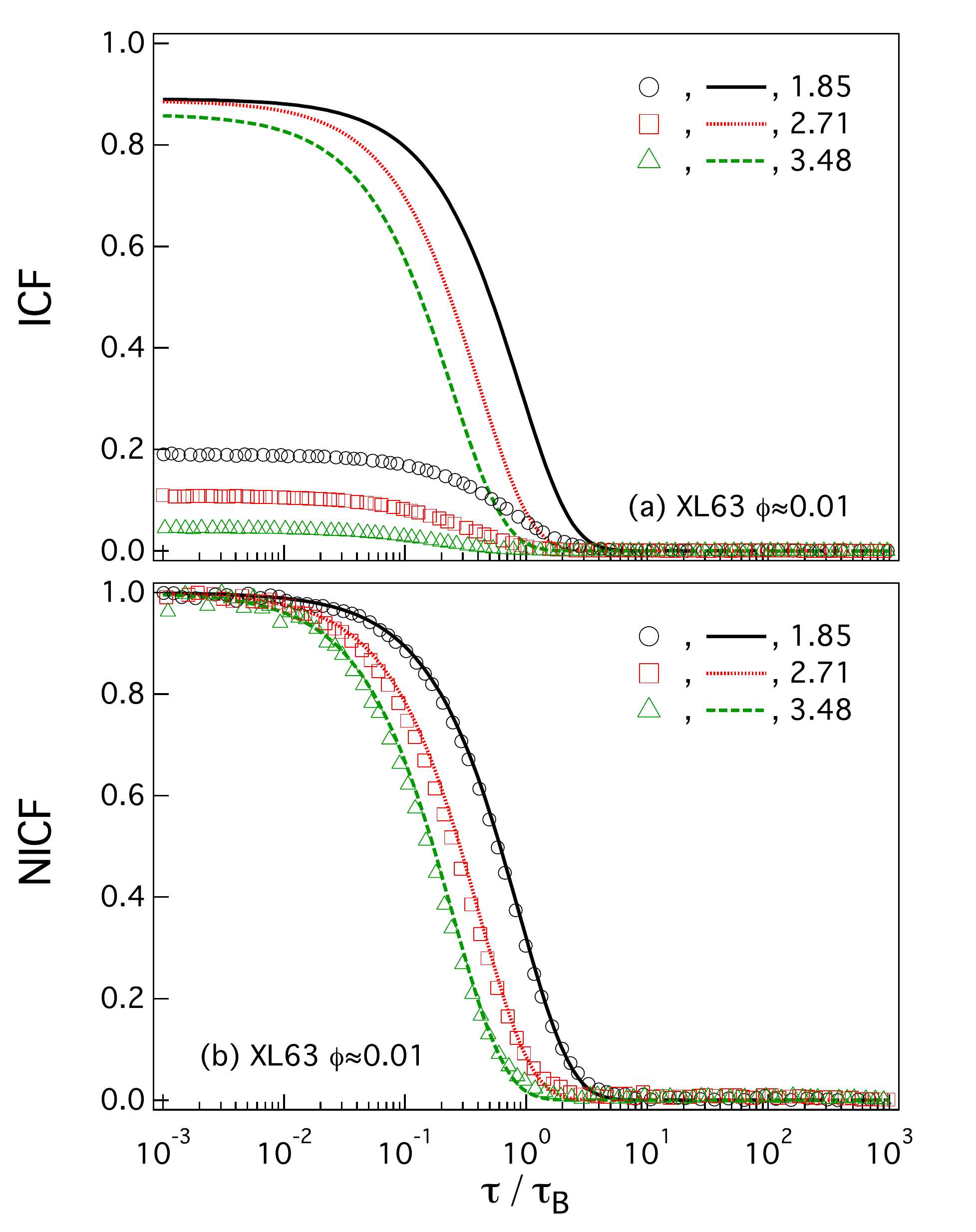}
\caption{XPCS (open symbols) and DLS (lines) measurements of the intensity autocorrelation functions for a dilute suspension $(\phi\sim0.01)$ of XL63 at the $qR$ values indicated; (a) un-normalized: $(g^{(2)}(q,\tau)-1)$  and; (b) normalized: $(g^{(2)}(q,\tau)-1)/c$.\label{Fig_1}}
\end{figure}

Fig. \ref{Fig_1} shows a comparison between the two techniques for a dilute suspension of hard sphere particles for three values of $qR$. Fig. \ref{Fig_1}a shows the ICF  (Eq. \ref{eq:Siegert}), which demonstrates the lower intercept of XPCS (symbols) relative to DLS (lines) - this is a consequence of the lower coherence of the X-rays relative to visible light. Fig. \ref{Fig_1}b shows $|f(q,\tau)|^2$, after normalization by the intercept, demonstrating that the two techniques give very similar results, despite the difference in intercept.

\begin{figure}
\includegraphics[scale=0.4]{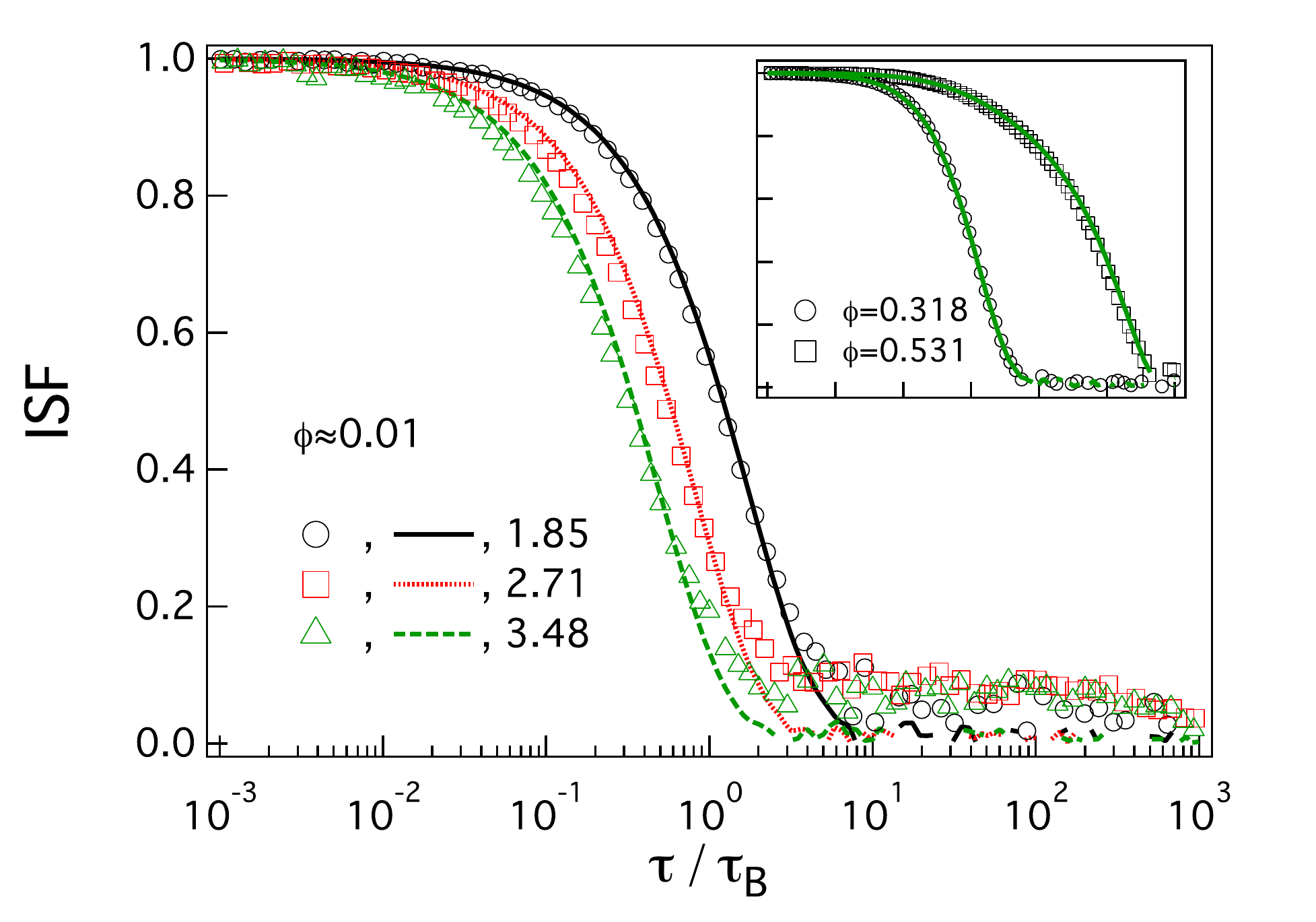}
\caption{Calculated ISFs from XPCS (open symbols) and DLS (lines) measurements on dilute suspensions at the $qR$ values indicated. The inset shows DLS measurements carried out in DLS cells (lines) and X-ray capillaries (open symbols) at the volume fractions indicated at a $qR$ value of 3.48 (near the peak of the structure factor).\label{Fig_2}}
\end{figure}

Fig. \ref{Fig_2} shows the ISFs for both techniques. As can be seen, the reduced intercept of the XPCS measurements now becomes important, and leads to the non-zero baseline for two of the scans. This is exacerbated by poor statistics, due to the fact that at low volume fractions the scattering is very low. This could be improved by accumulating for longer, however in the present case this was not attempted to limit the possibility of beam damage. Despite this difference, the results of the initial decay are still consistent.

DLS measurements were typically carried out in large sample cells with $\sim 8mm$ path lengths, whereas X-ray measurements must be carried out in small path length ($\sim1.5mm$) X-ray capillaries. To determine whether the use of capillaries causes any change in the dynamics (eg due to shear alignment during loading), the inset in fig. \ref{Fig_2} shows a comparison between DLS measurements at two volume fractions in standard DLS cells and in X-ray capillaries. The agreement seen here demonstrates that any difference observed between DLS and XPCS results is not due to different preparation methods or sample cells.

\begin{figure}
\includegraphics[scale=0.4]{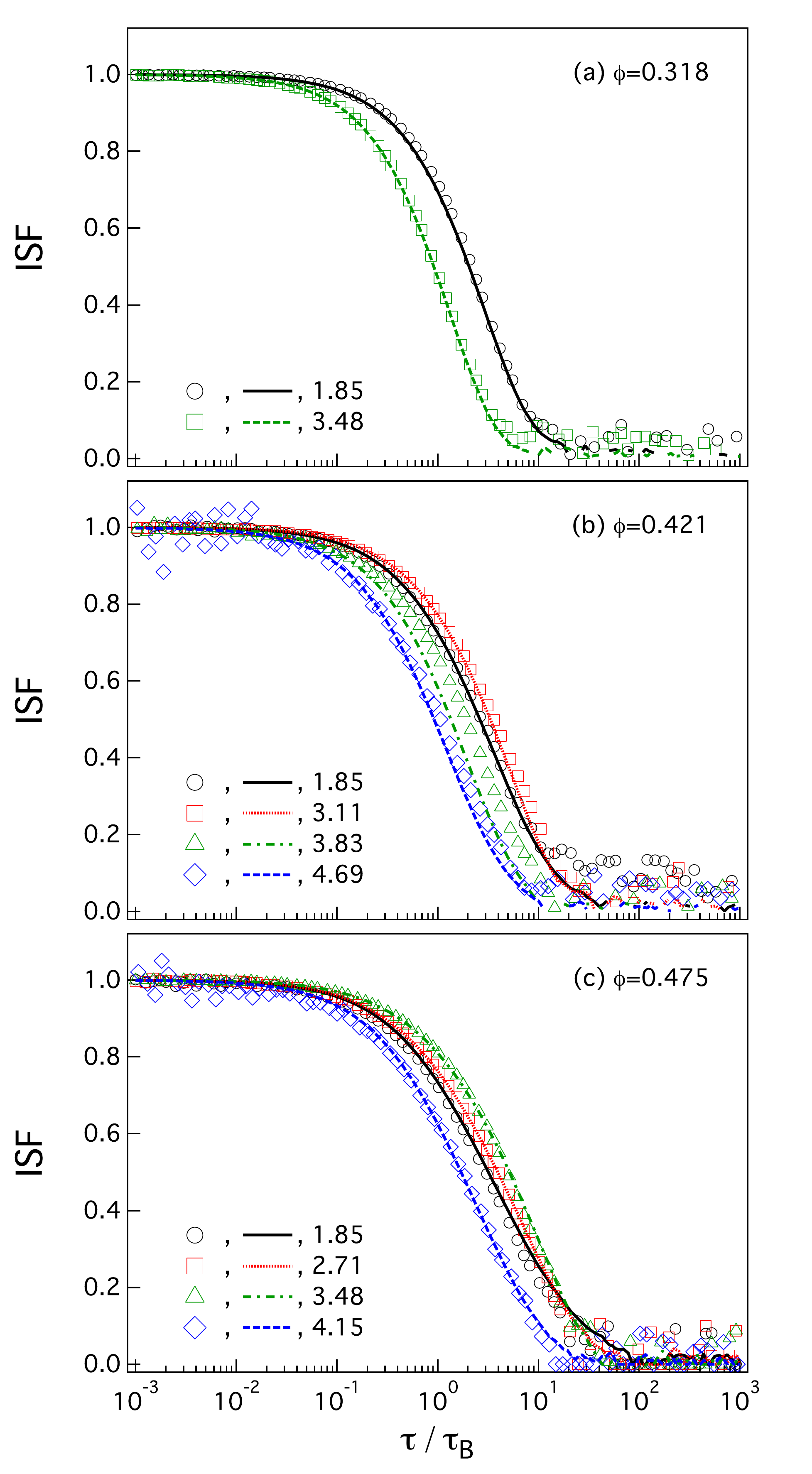}
\caption{ISF from XPCS (open symbols) and DLS (lines) at the volume fractions and $qR$ values indicated.\label{Fig_3}}
\end{figure}

Turning now to higher volume fractions, the measured ISFs from both techniques are shown in fig. \ref{Fig_3} for a range of volume fractions and scattering vectors. Fig. \ref{Fig_3} shows that, in general, there is reasonable agreement between the two techniques over the range of volume fractions shown, though deviations can be seen, particularly where the statistics are such that XPCS does not reach a zero baseline.

In order to compare the two techniques in more detail, we fit stretched exponential functions with the form $f(q,\tau)=\mathrm{exp}[-(\tau/\tau_{R}^{*})^{\beta})]$ by minimizing the Chi-square value using a form of non-linear least squares (Levenberg-Marquardt algorithm) with Igor Pro software from wavemetrics. DLS data are fit out to delay times $\tau/ \tau_B=20$. Due to the increased noise, XPCS data were fit over a smaller range, out to $\tau/\tau_B=5$.

\begin{figure}
\includegraphics[scale=0.4]{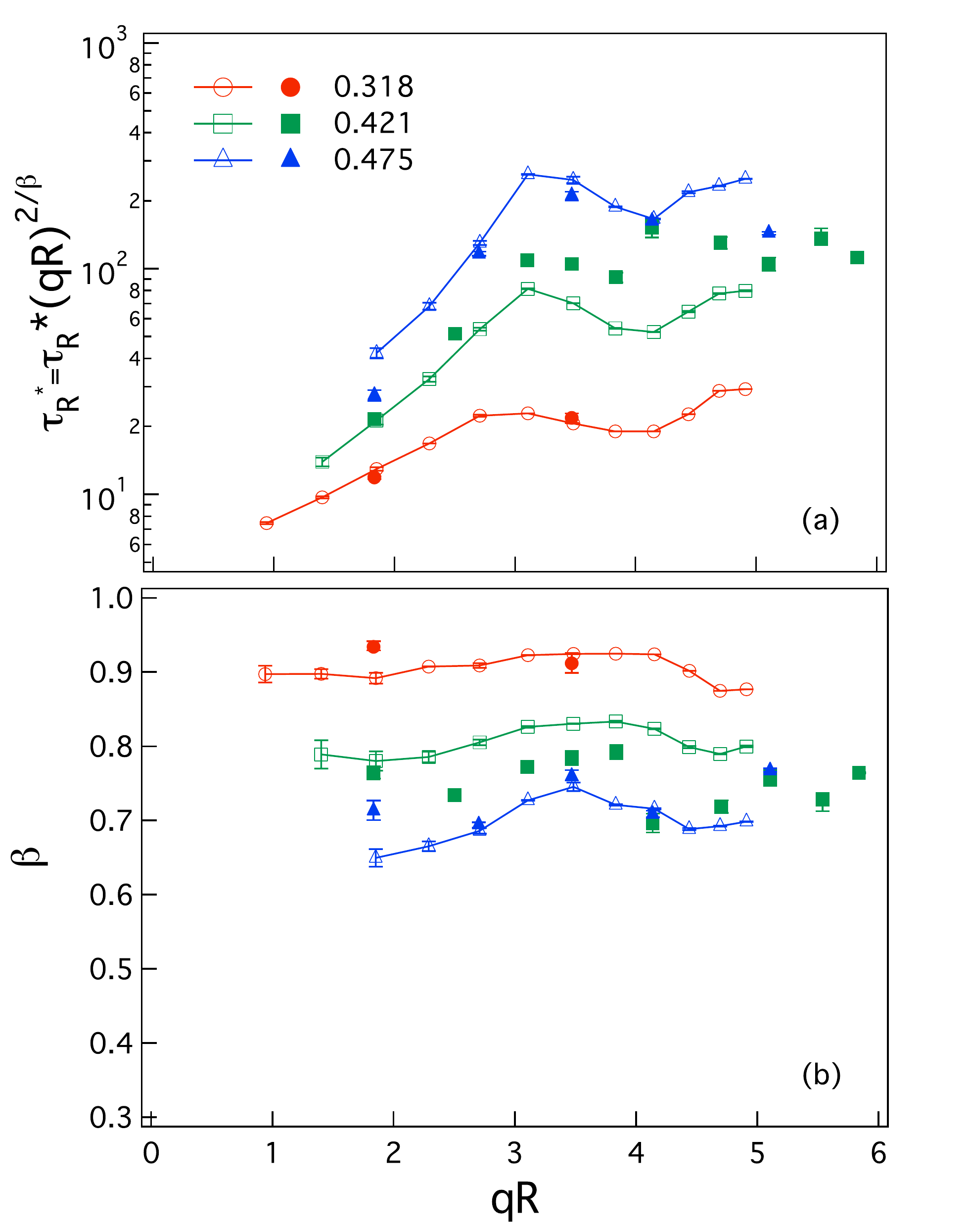}
\caption{(a) Characteristic time $\tau_{R}^{*}$ and (b) exponent $\beta$ of stretched exponential fits to the ISF, shown as functions of $qR$, at the volume fractions indicated. XPCS (filled symbols) and DLS (open symbols connected by lines). Error bars are estimated from the fits. For the characteristic times, the error in the Brownian time is also taken into account. Errors bars for DLS are smaller than symbols. Lines are drawn to guide the eye.\label{Fig_4}}
\end{figure}

The results of this analysis are shown in fig. \ref{Fig_4} for both DLS and XPCS data. Clearly there is agreement between the two techniques at volume fractions of 0.318 and 0.475. At a volume fraction of 0.421, XPCS gives characteristic times slightly larger than DLS and $\beta$ values slightly lower than DLS. However, the trends are clearly the same. The XPCS data, although it is subject to higher statistical noise, gives reasonable results out to higher $qR$ values. Most importantly, the two techniques show the same variation with $q$. Having established that XPCS and DLS data give consistent results we now turn to the question of scaling.

The above results show that the differences in scaling observed by Lurio et al. \cite{Lurio2000} and Segre and Pusey \cite{Segre1996} cannot be due to differences between the two techniques. Further, although the samples can become damaged if exposed to the X-ray beam for too long, these results show that, with care, damage can be avoided, suggesting that this is not an explanation for the differences observed in previous measurements. However, the possibility of beam damage provides an upper limit on how many runs can be made on each sample. By contrast, for DLS, an arbitrarily large number of runs can be made to achieve the desired statistical reliability. For the data studied here 50 runs of 1000s were routinely made. For this reason, having established the equivalence of the two techniques, the analysis below will be limited to the DLS data. 

\begin{figure}
\centering
\includegraphics[scale=0.4]{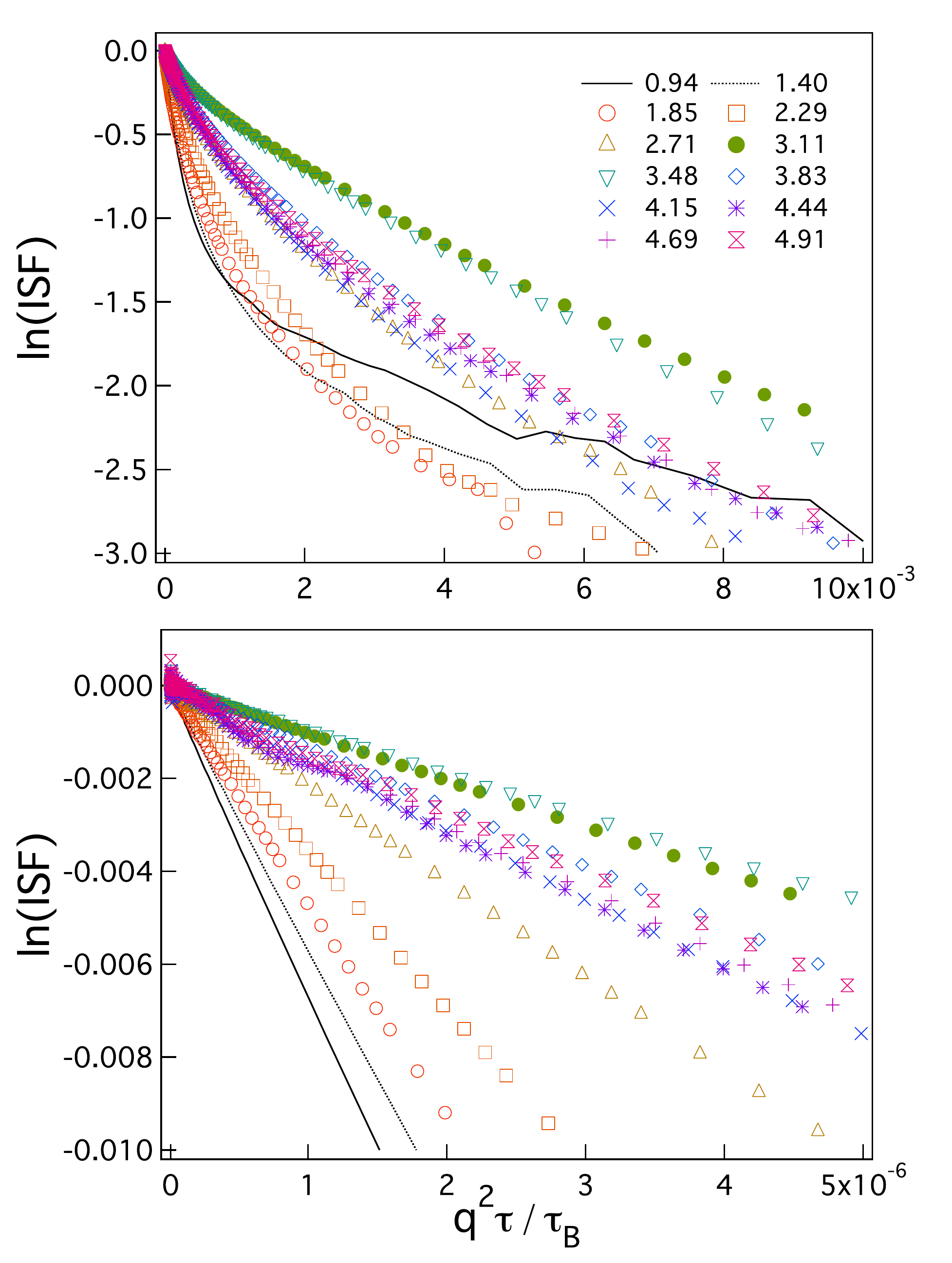}
\caption{Long time (top) and short time (bottom) behaviour of the ISF determined using DLS at $\phi=0.456$ for several $qR$ values as indicated. The representation is the one used in Segre et al \cite{Segre1996}. The linear behaviour at short time allows the identification of a short time diffusion coefficient $D_{s}(q)$.\label{Fig_5}}
\end{figure}

Fig. \ref{Fig_5} plots $\mathrm{ln(ISF)}$ vs $q^{2}\tau / \tau_B$ at a volume fraction of 0.456 for a range of scattering vectors for long (top) and short times (bottom). Note that the range of delay times shown is similar to those presented respectively in fig. \ref{Fig_1}a and fig. \ref{Fig_3}a of ref (1), and the noise on the DLS data is very low, demonstrating that our data are comparable with those of Segre et al.

\begin{figure}
\includegraphics[scale=0.4]{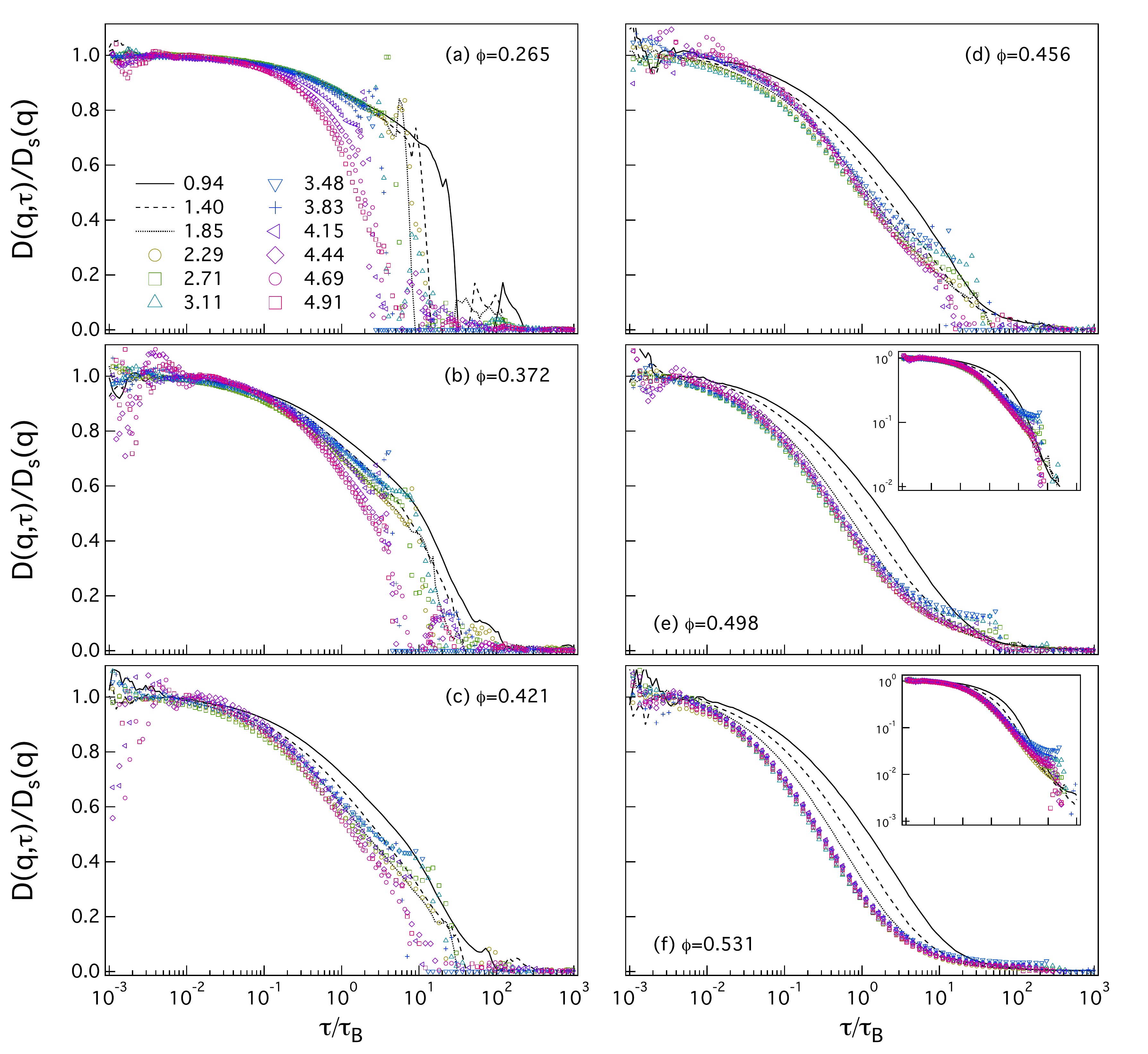}
\caption{Time-dependent diffusion coefficient $D(q,\tau)/D_{s}(q)$, from DLS experiments, normalised by the short-time diffusion coefficient at $qR$ values and volume fractions indicated. Insets show same data on a double-log scale. \label{Fig_6}}
\end{figure}

Segre et al. \cite{Segre1996} and then Lurio et al. \cite{Lurio2000} estimated short and long time diffusion coefficients by fitting straight lines to the initial and final decays of $\mathrm{ln(ISF)}$ vs $q^{2}\tau$,  as shown in Fig. \ref{Fig_5}. Such analyses inherently assume, from the outset, that the fastest and slowest detected density fluctuations are diffusive. The validity of this assumption will be discussed below. The full identification of the short and long time diffusive regimes requires the observation of plateaux in the time-dependent diffusion coefficient $D(q,\tau)$ (eq. 5) at short and long times respectively. To avoid duplicating unnecessary data, we present directly $D(q,\tau)/D_{s}(q)$, where $D_{s}(q)$ were obtained by identifying a plateau in $D(q,\tau)$ at short times. The results are shown in fig. \ref{Fig_6} for several scattering vectors at 6 volume fractions. The plateaux at short times suggest that the assumption of the existence of a short time diffusion coefficient is reasonable. However, this is not the case at long times, where there is no clear plateau in most cases, and thus the long time diffusion coefficient cannot be unambiguously determined.

Despite the noise, our results are in qualitative agreement with fig 2a of ref (1). A scaling of $D(q,\tau)/D_{s}(q)$ is reasonably observed at volume fraction $\phi \geq 0.456$ for $qR$ values spanning from 1.85 up to 4.91, even though noise is present at $\phi=0.456$. As the volume fraction is increased the $q$ dependence appears to decrease and the curves begin to converge. This trend is clearest for the volume fractions of 0.456, 0.498 and 0.531. At 0.456 the curves converge, within the noise, except for the lowest two $qR$ values of 0.94 and 1.40. At a volume fraction of 0.498, the data for $qR=1.85$ begins to deviate too, and at 0.531, in the metastable regime (ie where crystallization will occur at sufficiently long times), the deviation of the lowest three data sets becomes clearer. Interestingly however, while the deviation of the data for the lowest scattering vectors becomes more pronounced as volume fraction increases, the scaling for the scattering vectors around the peak becomes better. This scaling fails at long times, as highlighted in the log-log representation shown in the insets of fig. \ref{Fig_6} for the highest volume fractions. By contrast with the results presented here, Segre et al. clearly observed a plateau at long times at $\phi=0.456$ for all qR values spanning from 1.0 to 3.9 while we observe such plateaux only at qR values near the peak of the structure factor. Thus we question the existence of a long-time diffusive regime, and so a long-time diffusion coefficient $D_{L}(q)$, in our system.

\begin{figure}
\includegraphics[scale=0.4]{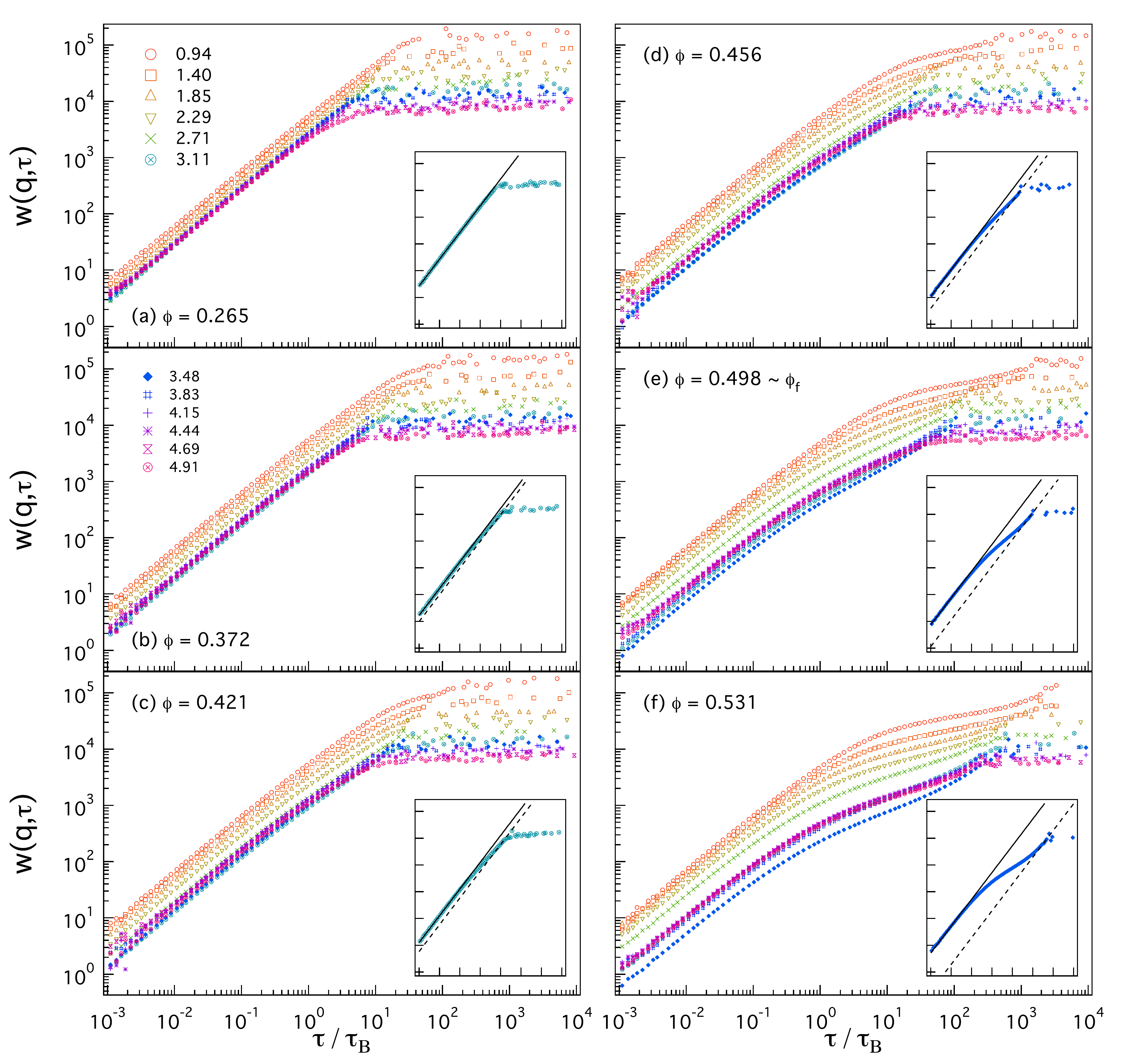}
\caption{Width function as function of delay time at $qR$ values and volume fractions indicated. Lines and dashed lines are linear fits of $log[w(q,\tau)]=log(D_{s}(q))+log(\tau)$ at, respectively, short and long times. Inset shows the width function at the $qR$ values nearest to the peak of structure factor for the corresponding volume fraction (scale as main panel).\label{Fig_7}}
\end{figure}

\begin{figure}
\includegraphics[scale=0.4]{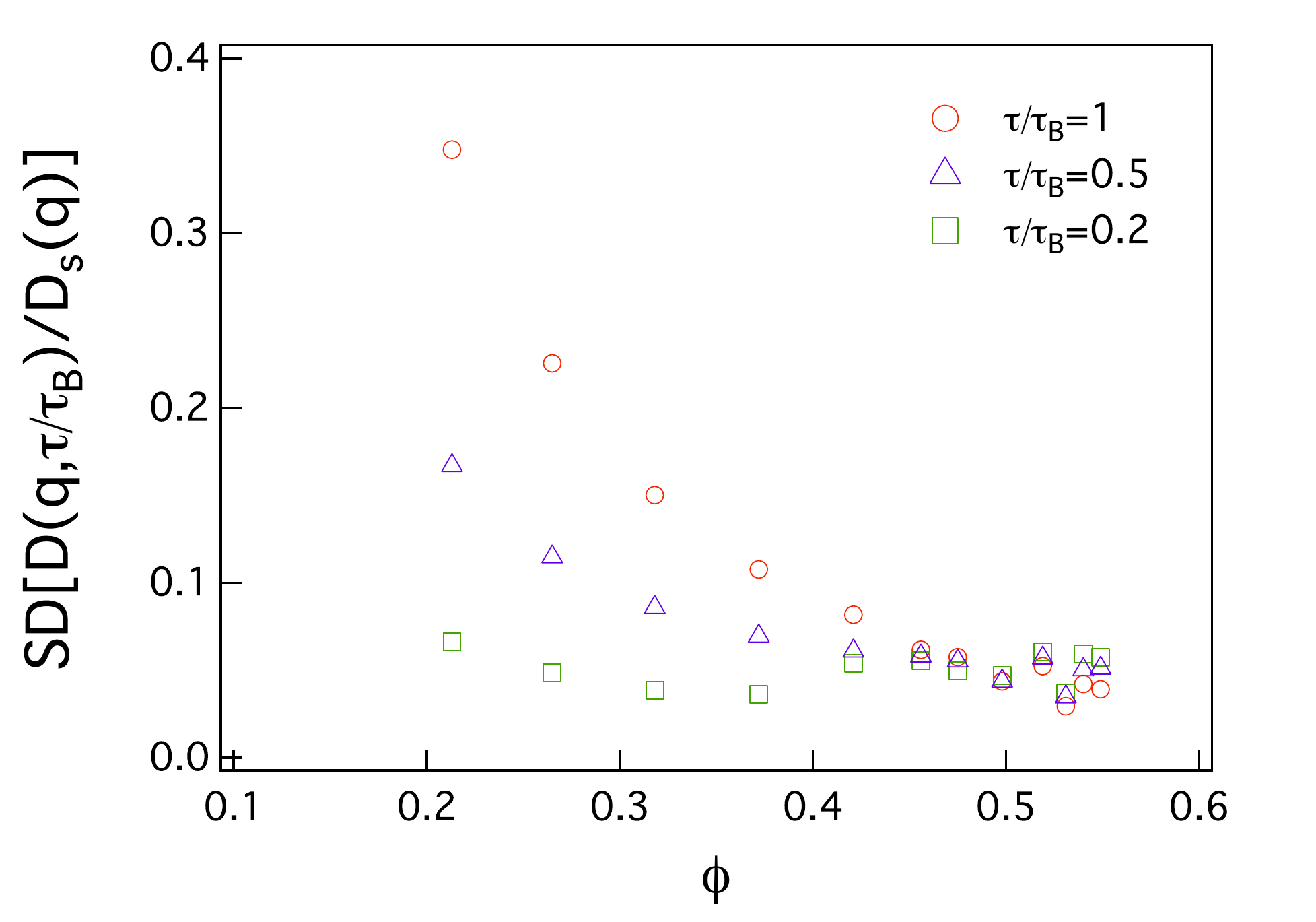}
\caption{Standard deviation of $D(q,\tau/\tau_{B})/D_{s}(q)$, from DLS experiments, at fixed values of $\tau/\tau_{B}$ and for $qR$ values in the range [2.29,4.91].\label{Fig_8}}
\end{figure}

\begin{figure}
\includegraphics[scale=0.4]{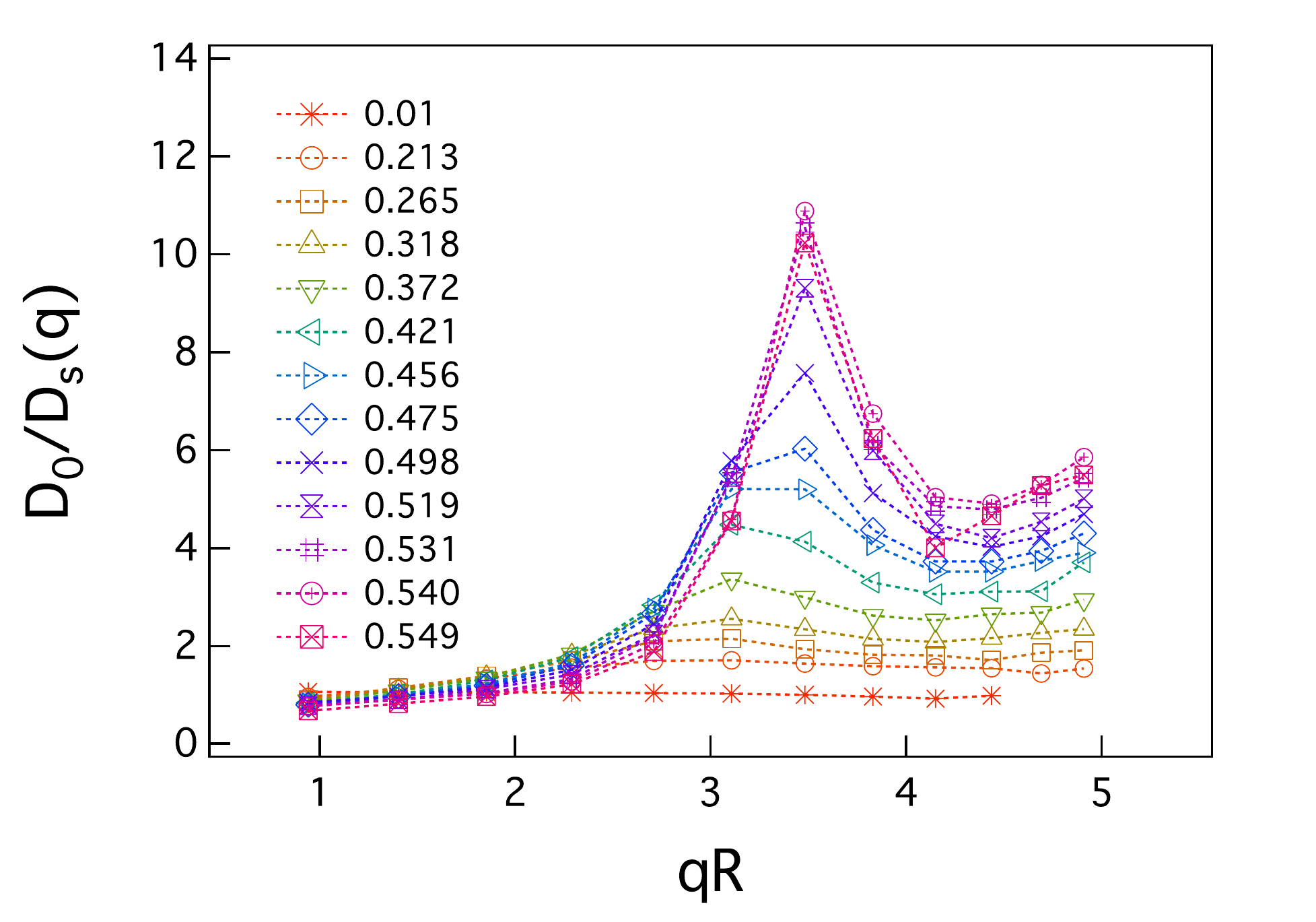}
\caption{Inverse of the short time diffusion coefficient expressed in term of the free diffusion coefficient $D_{0}$ at volume fractions indicated. Lines are drawn to guide the eye.\label{Fig_9}}
\end{figure}

\begin{figure}
\includegraphics[scale=0.4]{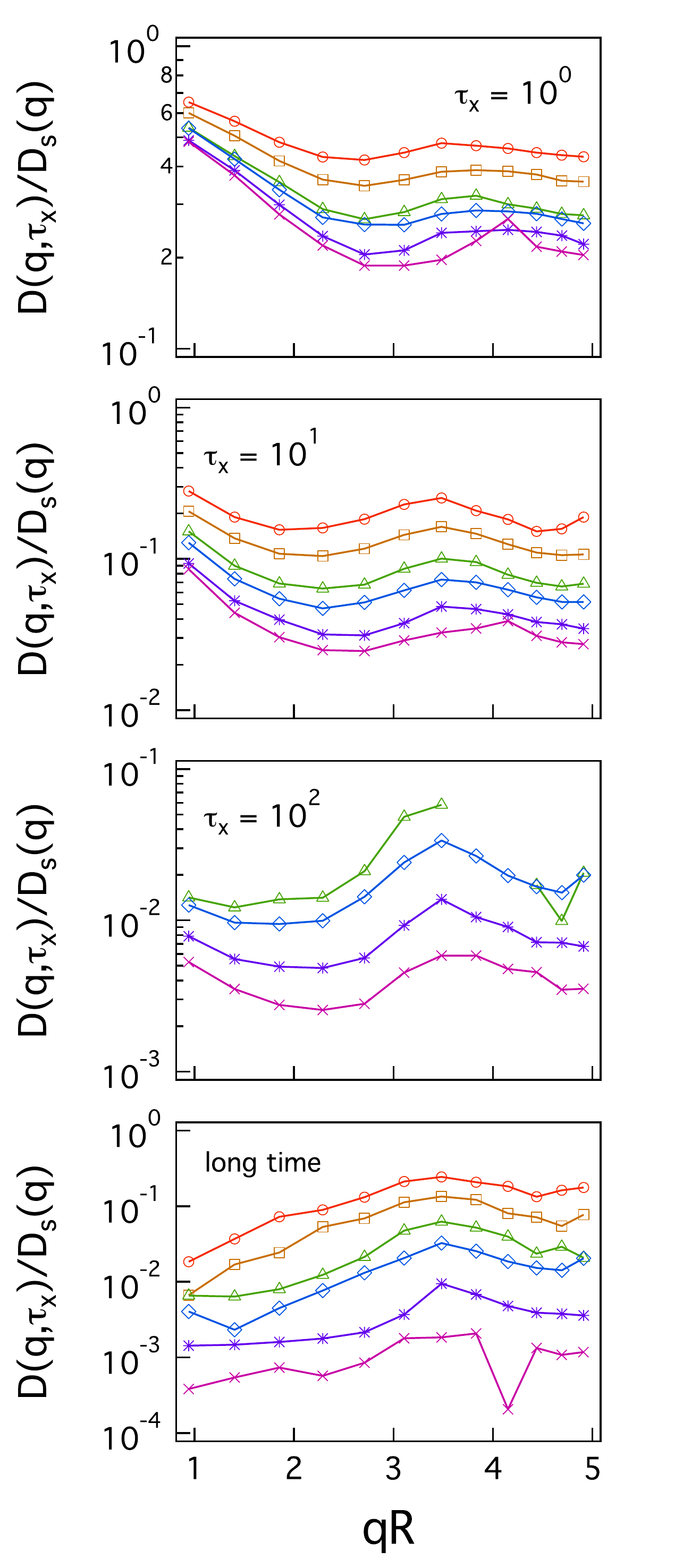}
\caption{Ratio of $D(q,\tau_x)/D_s (q)$ as a function of $qR$ at volume fraction $\phi$ equal to 0.475 (circles), 0.498 (squares), 0.519 (triangles), 0.531 (diamonds), 0.540 (stars) and 0.549 (crosses). Each graph corresponds to a specific $\tau_{x}$ value as indicated. The bottom panel shows $D(q,\tau_x)/D_s (q)$ at the longest time before noise starts to dominate. Note that this time increases with volume fraction. Within the noise there is no real indication of scaling between short and long time behaviour.\label{Fig_10}}
\end{figure}

\begin{figure}
\includegraphics[scale=0.4]{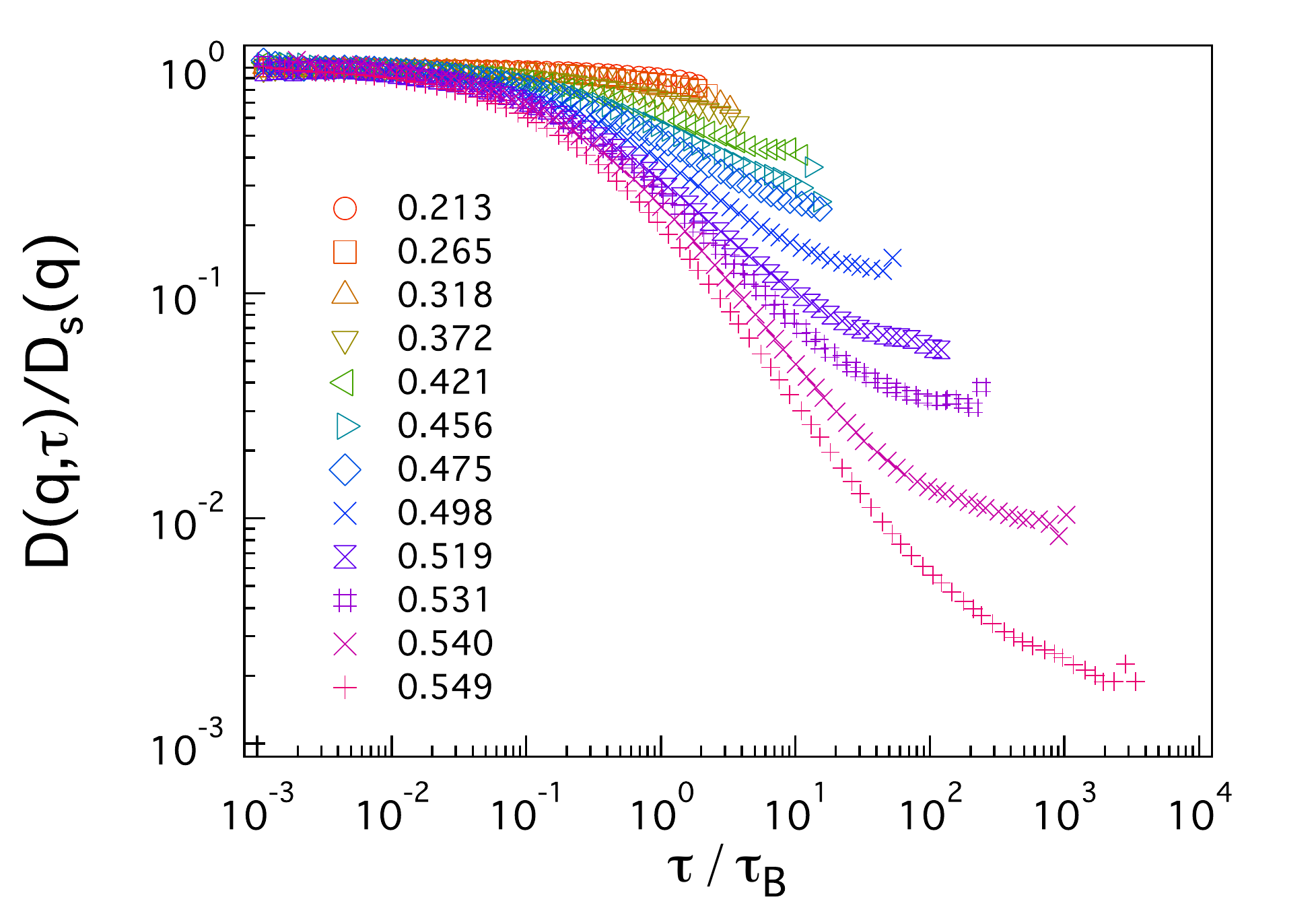}
\caption{Time-dependent diffusion coefficient, normalised to the short-time diffusion coefficient, as a function of delay time for several volume fractions at $qR=3.57$ (around the peak of the structure factor for higher volume fractions).\label{Fig_11}}
\end{figure}

The numerical derivative required to calculate $D(q,\tau)$, introduces noise and makes it difficult to identify a long time diffusive regime, where such a regime exists. An alternative analysis which is not subject to this noise, and more clearly exposes the diffusive regimes, is to calculate the width function \cite{Martinez2008}, defined by analogy with the mean square displacement as:
\begin{equation}
w(q,\tau)=-\mathrm{ln}[f(q,\tau)]/q^{2}
\end{equation}
which is shown as a function of delay time in fig. \ref{Fig_7}. The insets show the width function at $q_m$. In this representation diffusive regimes are identified where $w(q,t)$ grows linearly with delay time. This determination of diffusion does not require numerical differentiation of the data (in contrast to fig. \ref{Fig_6}), and is insensitive to the time scale on which the ISF is plotted (in contrast to fig. \ref{Fig_5}). Clearly, this representation shows unequivocally that short time diffusive regimes are observed over several decades for all volume fraction and $qR$ values probed. However fig. \ref{Fig_7} confirms that difficulties occur in identifying linear long-time behaviour at some $q$ values. In other words, the ISF and so the width function attain the experimental noise floor before the anticipated long-time diffusive regime is observed. For the data shown here, long-time diffusive behaviour can only be discerned at $q_m$ for $\phi \geq 0.498$, and this is shown in the insets - the long time diffusive regime is not attained for other $q$ values or for lower volume fractions. This is consistent with recent experiments which showed that the mean-squared distance particles must traverse, in order for density fluctuations to forget the effects of packing constraints, is smallest at $q_m$ \cite{Megen2009}. A consequence of this is that we are unable to directly conclude whether or not the ratio $D_{s}/D_{L}$ is independent of $q$, as found by Segre et al. \cite{Segre1996}, a result which was disputed by Lurio et al. \cite{Lurio2000}.
Returning to the scaling behaviour, as $\phi$ is increased, $D(q,\tau)/D_{s}(q)$ seems to scale better in the first decades even though the last decades shows an increasing q-dependence once $\phi \geq 0.498$. To quantify this observation, we choose three fixed delay times, and plot in fig. \ref{Fig_8} the standard deviation of the ÒspreadÓ of $D(q,\tau)/D_{s}(q)$ values at these times as a function of volume fraction. The scaling is clearly observed via the convergence of this quantity as volume fraction is increased. Of course the decay of the ISF shifts to longer delay times as volume fraction increases, so this analysis is limited. However it clearly demonstrates the scaling as volume fraction increases.
Keeping in mind that the determination of the plateau at long times is ambiguous, particularly at low volume fractions, fig. \ref{Fig_9} shows the $qR$ dependence of the inverse of the short time collective diffusion coefficient. Clearly $D_o/D_s$ scales with the structure factor in agreement with previous results eg Segre and Pusey \cite{Segre1996}.
We stated above that the determination of the long time diffusion coefficient is problematic. In order to remove the ambiguity associated with the measurement of a long time plateau, we simply determine the apparent diffusion coefficients $D(q,\tau_x)$ at several delay times $\tau_x$. The ratio $D(q,\tau_x)/D_{s}(q)$ is then plotted in fig. \ref{Fig_10} for several volume fractions. At the lowest volume fraction measured there is some hint of scaling only at the early times but clearly not at long times. As the volume fraction increases any hint of scaling disappears, and the ratio never demonstrates q-independence at any delay time.
Finally, Fig. \ref{Fig_11} shows $D(q,\tau)/D_{s}(q)$ near the peak in the structure factor for several volume fractions. This graph confirms that at the peak it is just about possible to define a diffusive regime at long times (see also fig. \ref{Fig_7}f). Of more interest is the volume fraction dependence of the rate at which the behaviour deviates from diffusive. At $\tau/\tau_{B}=10^0$(after three decades in time) $D(q,\tau)/D_{s}(q)$ has dropped by almost an order of magnitude at a volume fraction of 0.549,  whereas $D(q,\tau)/D_{s}(q)$ remains approximately 1 at a volume fraction of 0.213. So, as the volume fraction increases, the system deviates more quickly from diffusive behaviour, even though the dynamics are slower.

\section{Discussion}
The first part of the results compared DLS and XPCS results on identical samples. As far as we are aware, this is the first comprehensive comparison between DLS and XPCS on concentrated hard sphere samples. The results demonstrate a number of things. Within statistical errors, the ISFs determined from DLS and XPCS agree very well. The main limitation of XPCS is that good sample statistics are much harder to obtain, partly due to the lower coherence, and partly due to beam damage, which limits how long statistics can be accumulated for on a particular sample. This is particularly true for the higher volume fractions, where restricted diffusion means individual particles are exposed to the X-ray beam for long periods of time. In order to improve the statistics at high volume fractions, it would be desirable to have the samples mounted on a translation stage so that a series of short measurements can be made at different positions in the sample to limit sample damage. Such studies are currently being planned.

Despite this, the results confirm that careful use of XPCS can provide data of similar quality to DLS. There are a number of situations where XPCS would be preferable to DLS: first, where particles are smaller than $R\sim 120 nm$, where the structure factor peak cannot be accessed using DLS; second, for samples where multiple scattering is significant; and finally, where dynamics need to be studied at higher $q$ values as shown in fig. \ref{Fig_4} (though not explored here, even higher values are possible using XPCS).

We turn now to the results of the scaling analysis. First, we find that $D_{s}(q)$ scales with $S(q)$, an effect knows as DeGennes narrowing. Second, we find that the long time diffusion coefficient cannot be defined away from the peak for our system, so the scaling of the ratio $D_{s}/D_{L}$ observed by Segre and Pusey \cite{Segre1996}, but not by Lurio et al. \cite{Lurio2000}, cannot be confirmed. Third, we do observe a scaling behaviour of the time-dependent diffusion coefficient at high volume fraction. This result agrees with Segre and Pusey but is not observed by Lurio et al.

As the system under study is a hard sphere system very similar to the system used by Segre and Pusey, (though with higher polydispersity) the difference in the results is puzzling. One possible explanation is in the way the long time diffusion coefficient is determined. Our results have shown that the determination of this quantity is problematic, and according to our analysis, it cannot be defined away from the structure factor peak. Inspection of Fig. \ref{Fig_11}, for example, shows that a clear plateau can be identified at early times ($D_s$), but that a plateau is much less well defined at long times. It is possible that the differences between our results and those of Segre and Pusey are related to the determination of this property.

The identification of a long-time diffusive regime for the collective dynamics (through the width function, fig. \ref{Fig_7}) is clearly observed around $q_m$ but not at other $q$ values. So the following question can be asked: does a long-time diffusive regime exist for all $q$? We consider the quantity $\langle\Delta r^{2}(\tau_{m}^{c}(q))\rangle$, which corresponds to the mean squared distance a particle has to move for the number density fluctuations to forget excluded volume effects - where $\langle\Delta r^{2}\rangle$ and $\tau_{m}^{c}(q)$ are respectively the time-dependent mean square displacement and the delay time at the crossover between the fast and slow processes. This quantity was previously introduced and measured for the same particle suspensions \cite{Megen2009}. At $\phi_f=0.498$ and $q_m$, the quantity $\langle\Delta r^{2}(\tau_{m}^{c}(q_{m}))\rangle$ is smaller by a factor of $\sim5$ compared to $\langle\Delta r^{2}(\tau_{m}^{c}(q))\rangle$ at qR=1.0. This quantifies the difficulty in identifying the long-time diffusive behaviour from the coherent ISF at $q$ vectors away from the peak of the structure factor, simply because particles are unable to move the distances necessary for the number density fluctuations to forget packing constraint effects.

Interestingly, this question has been independently addressed in a recent paper by Holmqvist and Nagele \cite{Holmqvist2010} for charge stabilized colloids. These authors do observe the long time scaling behaviour found by Segre and Pusey \cite{Segre1996} for scattering vectors around the peak of the structure factor. One difference between their system and the one studied here is that due to the smaller size, lower viscosity and perhaps the nature of the interactions, the timescales are very much faster - the ISF is complete within $\approx30$ ms, whereas for the data presented here the decays at the highest volume fractions are of the order of 10 s, almost 1000 times longer. In our experiments 50 x 1000 s measurements were made, but we were still unable to observe long time diffusive behaviour except at the peak of the structure factor. So it may be that for the hard sphere system used here, the determination of a true long time limit away from the structure factor peak is not possible with any reasonable experimental time. Nevertheless, an analysis similar to our figure 7 for the data shown in Holmqvist and Nagele \cite{Holmqvist2010} would provide a more unambiguous measure of the determination of the long time diffusion coefficient, and would shed more light on the observed differences.

More generally, the results summarized in fig. \ref{Fig_11} show that the evolution of the time-dependent diffusion coefficient changes dramatically with volume fraction. At low volume fractions the transition between the two takes a long time. However as volume fraction increases the short and long time processes quickly separate, as more and more particles become trapped in neighbour cages. The magnitude of this separation increases, and occurs at shorter times, as the dynamics become slower.

\section{Conclusions}
We present an extensive study of concentrated colloidal hard sphere suspensions using both DLS and XPCS, as functions of volume fraction and scattering vector. A scaling behaviour, found by Segre et al. \cite{Segre1996}, is observed over several decades in time but not in the long time regime where significant q-dependence is observed for higher volume fractions. While the short-time diffusion regimes clearly exist at all volume fractions and scattering vectors, long-time diffusive regimes are only observed near the peak of the structure factor at high volume fractions. Thus the existence of a collective long-time diffusive regime is rather questionable, even though an ÒapparentÓ linear regime is observed in the common representation $\mathrm{ln(ISF)}$ vs $q^{2}\tau$ used by both Lurio et al \cite{Lurio2000}. and Segre and Pusey \cite{Segre1996}. By interpreting the quantity $\langle\Delta r^{2}(\tau_{m}^{c}(q))\rangle$, measured by van Megen et al. \cite{Megen2009}, we explain the absence of a long-time diffusive regime by simply arguing that particles are unable to move the distances necessary for the number density fluctuations to forget packing constraint effects during accessible time scales.

\begin{acknowledgments}
We acknowledge financial support from the \textit{Access to Major Research Facilities Programme} which is a component of the \textit{International Science Linkages Programme} established under the Australian Government's innovation statement, \textit{Backing AustraliaÕs Ability}. We acknowledge the European Synchrotron Radiation Facility for provision of synchrotron radiation.\end{acknowledgments}


%
%

%



\end{document}